\documentclass[journal]{IEEEtran}

\ifCLASSINFOpdf
\else
   \usepackage[dvips]{graphicx}
\fi
\usepackage{url}

\usepackage{booktabs}
\usepackage{graphicx}
\usepackage{multirow}
\usepackage{amsfonts}

\begin{document}

\title{Detecting Replay Attacks Using Multi-Channel Audio: A Neural Network-Based Method}
\author{Yuan Gong,~\IEEEmembership{Student Member,~IEEE,}
        Jian Yang,
        Christian Poellabauer,~\IEEEmembership{Senior Member,~IEEE}% <-this % stops a space
\thanks{The authors are with the Department of Computer Science and Engineering, University of Notre Dame, IN, 46637 USA (e-mail:ygong1@nd.edu).}}
\markboth{}
{Shell \MakeLowercase{\textit{et al.}}: Bare Demo of IEEEtran.cls for IEEE Journals}

\maketitle

\begin{abstract}

With the rapidly growing number of security-sensitive systems that use voice as the primary input, it becomes increasingly important to address these systems' potential vulnerability to replay attacks. Previous efforts to address this concern have focused primarily on single-channel audio. In this paper, we introduce a novel neural network-based replay attack detection model that further leverages spatial information of multi-channel audio and is able to significantly improve the replay attack detection performance.
\end{abstract}


\IEEEpeerreviewmaketitle

\section{Introduction}

In recent years, an increasing number of security-sensitive voice-controlled systems such as virtual assistants have been introduced. While these systems are typically equipped with a speaker verification model, their vulnerabilities to multiple types of replay attacks have become a new security concern, e.g., an attacker can play a pre-recorded or synthesized speech sample to spoof the speaker verification system~\cite{08diao2014your,lei2017insecurity,03carlini2016hidden,05gong2017crafting,ijcai2019-649}. Therefore, developing an effective countermeasure to distinguish between genuine and replayed samples has become a recent research focus~\cite{kinnunen2017asvspoof,todisco2019asvspoof,kamble2020advances,gong2018overview}. While there have been many prior efforts in this area~\cite{jelil2017spoof,witkowski2017audio,02todisco2017constant,bakar2018replay,lavrentyeva2017audio,li2017study,cai2017countermeasures,wang2017feature,chen2017resnet,Jelil2018,kamble2018effectiveness,Yang2018,gong2018protecting,Patil2019}), they only focus on detecting replay attacks based on \emph{single-channel} input and therefore only leverage the temporal and spectral features. However, we identified three reasons why a countermeasure designed using \emph{multi-channel} audio input could provide improved performance. First, multi-channel audio captured by a microphone array contains spatial information, which can contain useful cues to help distinguish genuine and replayed samples~\cite{gong2019remasc,zhang2016voicelive}. Second, it is relatively easy for an attacker to manipulate the temporal and spectral features to fool an anti-spoofing module~\cite{liu2019adversarial} by simply modifying the replayed signal, while spatial features (e.g., time difference of arrival (TDoA)) are harder to manipulate and hence are more reliable. Third, multi-channel speech recognition techniques have been extensively studied and adopted~\cite{benesty2008microphone,sainath2017multichannel} and most modern far-field speech recognition systems are already equipped with microphone arrays, which makes it easy to obtain multi-channel audio. 

Only few prior studies focused on multi-channel voice anti-spoofing. In~\cite{zhang2016voicelive}, the authors propose VoiceLive, which captures TDoA changes in a sequence of phoneme sounds to the two microphones and uses such unique TDoA dynamics to distinguish between replayed and genuine samples. The limitation is that this method requires the microphones to be placed very close (1-6cm) to the mouth. In~\cite{shiota2015voice}, the authors use the ``pop noise'' caused by breathing to identify a live speaker based on two-channel input, where one channel is used to filter the pop noise and another is used as a reference. The limitation is that the pop noise effect disappears over larger distances and thus the method is only applicable to close-field situations. In~\cite{ryoyaimproving,yaguchi2019replay}, the authors use generalized cross-correlation (GCC) of the non-speech sections of a stereo signal to detect the replay attack. The idea is that loudspeakers tend to generate electromagnetic noise during the non-speech section, and therefore the cross-correlation between the two-channel signals will be higher for replayed samples than genuine samples in the non-speech section. The limitation is that in order to make the electromagnetic noise detectable, a suitable background noise level and high-fidelity microphone is required, which is not always met in realistic settings. 

In summary, previous multi-channel replay attack detection methods: 1) have only been designed for two-channel audio input while modern microphone arrays usually have more microphones and contain richer spatial information; 2) rely on hand-crafted features and calibrating the features for a different microphone array system or a different environment can be difficult; and 3) have relatively few applicable scenarios due to the close-field or low SNR requirement. 

In order to overcome the above-mentioned limitations, in this paper, we propose a novel neural network-based replay attack detection model that has the following advantages. First, the proposed model is completely data-driven, i.e., no manual spectral or spatial feature engineering is needed, and the model can be used for inputs of any number of channels without knowing microphone array specifics (such as the array geometry) whenever training data is available. For the same reason, the proposed model can adapt to different environments using the training data and, therefore, there are no explicit constraints on usage scenarios. Second, all components (i.e., beamformer, feature extraction, and classification) are part of a neural network framework, which makes it easy to train them using existing massive optimizing methods and combine them with other neural-based countermeasure models. This work is the first neural-based multi-channel replay attack detector. We perform experiments using the recently collected ReMASC corpus~\cite{gong2019remasc} that contains genuine and replayed samples recorded in a variety of environments using different microphone array systems. We find that by leveraging the multi-channel audio input, a significant performance improvement (up to 34.9\%) can be achieved compared to a single-channel input model.

%--------------------------------------%
\section{The Multi-channel End-to-end Replay Attack Detection Network}
\label{sec:method}

One classic microphone array signal processing technique is the filter-and-sum beamformer \cite{benesty2008microphone}. For a multi-channel audio with $C$ channels $x=(x_1, x_2, ..., x_C)$, the filter-and-sum beamformer filters each audio channel $x_c$ using an $N$-point FIR filter $h_c$ with a delay (or advance) of a steering time difference of arrival (TDOA) $\tau_c$ to conduct the time alignment, and then sum the output of each channel together to obtain the output $y$:
\begin{equation}
\label{equ:1}
    y[t]=\sum_{c=1}^{C}\sum_{n=1}^{N}h_c[n]x_c[t-n-\tau_c]
\end{equation}

The filter-and-sum beamformer can be decomposed into two sub-processes: 1) estimating the $\tau_c$ and 2) finding the optimal filter $h$. The first sub-process can be done by using a separate time-delay estimation module. But in order to implement the entire model in a neural network framework, we follow the method described in~\cite{sainath2017multichannel} to implicitly absorb the steering delay into the filter parameters and use a bank of $P$ filters for each channel to capture different $\tau_c$:
\begin{equation}
\label{equ:2}
    y[t] = (y^1[t], y^2[t], ..., y^P[t])
\end{equation}
where:
\begin{equation}
\label{equ:3}
    y^p[t]=\sum_{c=1}^{C}\sum_{n=1}^{N}h_c^p[n]x_c[t-n]=\sum_{c=1}^{C}x_c[t]\ast h_c^p
\end{equation}

and where $\ast$ denotes the convolution operation. In the remainder of this paper, we refer to $h$ as a \emph{front-end filter} to distinguish it from other filters in the neural network.

Usually, an optimal filter $h$ is designed using a separate optimization objective (e.g., minimum variance distortionless response (MVDR)~\cite{van1988beamforming} or multichannel Wiener filtering (MWF)~\cite{brandstein2013microphone}), which is usually different from the objective of the actual learning task (e.g., word error rate for the speech recognition task). While this is acceptable for tasks such as speech recognition since low speech distortion and noise level are likely to improve the recognition accuracy, it might lead to the opposite effect for the replay attack detection task, because the filters might remove the useful cues in noisy or high-frequency components that the replay attack detector relies on. Therefore, a better strategy is to jointly optimize the beamformer with the replay attack detector. 

\begin{figure}[t]
  \centering
  \includegraphics[width=6.42cm]{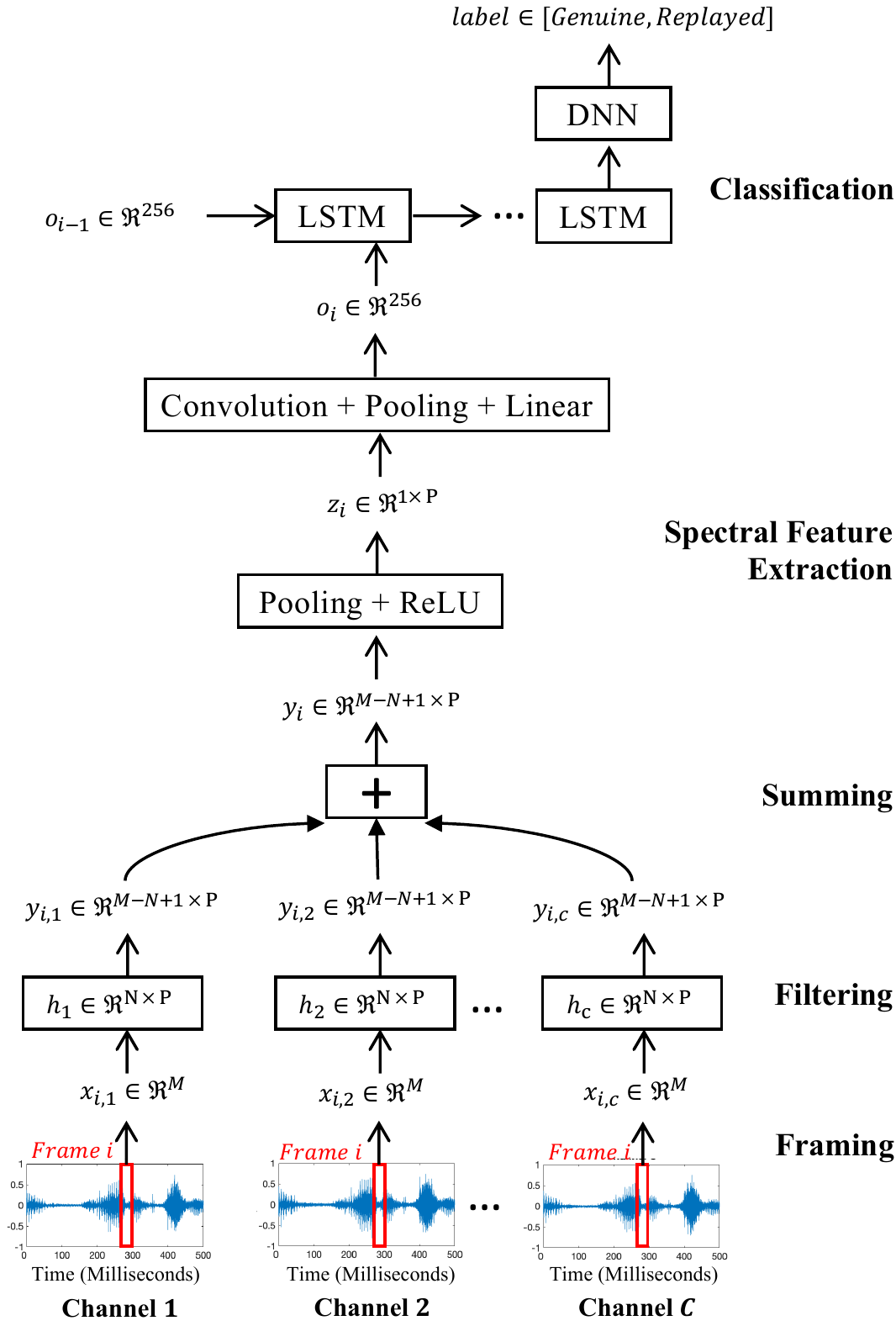}
  \caption{Design of the multi-channel replay attack detection network.}
  \label{fig:nn}
\end{figure}

We design the network based on the architecture presented in~\cite{sainath2017multichannel,sainath2015convolutional}, which was previously used for the speech recognition task. As shown in Fig.~\ref{fig:nn}, first, we perform a non-overlapped framing of 20ms to the input waveform (frame length $M$=882 at a sample rate of 44.1kHz) and feed each frame $x$ to the next layer. Then, we conduct the filter-and-sum beamforming as described in Equations~\ref{equ:2} and~\ref{equ:3} to obtain a 2-D representation $y$ for each frame. Note that since the front-end filters also conduct frequency decomposition in addition to spatial filtering, $y$ can be viewed as a time-frequency representation~\cite{sainath2017multichannel,sainath2015learning}. We use filter length $N$=630 and test different values for the filter number $P$ in our experiments. The ratio of $N$ to $M$ is the same as that in~\cite{sainath2017multichannel}. Note that the filter-and-sum beamformer differs from the simpler delay-and-sum beamformer in that an \emph{independent} weight is applied to each of the channels before summing them. Therefore, the filters $h$ for each channel do not share the weight. Similar convolution layer designs have been widely adopted in image processing tasks to process the three-channel RGB input, but not for the purpose of beamforming. To lower the computational overhead, we do not apply padding for this convolution layer. After that, we conduct a global max-pooling in time and apply a ReLU~\cite{nair2010rectified} nonlinear activation function to the beamformer output and get $z\in\mathfrak{R}^{1\times P}$ and feed it to a frequency convolution layer that consists of 256 1$\times$8 filters with a max-pooling of size 3. The pooled output is then fed to a 256-dimensional fully-connected layer. The output $o\in\mathfrak{R}^{256}$ is the representation of this frame. We conduct the same above operations to all frames and feed the sequence of $o$ to three stacked LSTM~\cite{hochreiter1997long} layers, each with 832 hidden units and feed the output of the last frame to a single fully-connected layer to obtain the prediction. The entire model is end-to-end from audio waveform to prediction and is trained jointly using a unified objective of minimizing the weighted cross-entropy loss. We re-weight the cross-entropy loss for each class using the normalized reciprocal of the sample number of the class in the training set to avoid the class-imbalance problem. 

Since the replay attack detection is a sample-level classification task, i.e., there is only one label for each audio sample (either genuine or replayed), using part of the input can be sufficient and will significantly lower the computational overhead and avoid potential overfitting. In this work, we use the beginning 1s of each audio sample as the input to the network. Compared to using a segment from the middle of a sample, the beginning part of a sample usually contains a mixture of non-speech and speech samples, which can be beneficial for the replay attack detection task, which we verify in our experiments shown in~\ref{sec:exp3}.

To summarize, the proposed model has the following advantages: 1) It is completely data-driven, hence no manual feature engineering is needed, and the model can be used for inputs of any number of channels without knowing the microphone array information such as array geometry whenever training data is available. 2) All components (beamformer, feature extraction, and classification) are in the neural network framework, which makes it easy to train using existing massive optimizing methods. The intermediate tensor $y$ is a standard time-frequency representation, so it is easy to further improve the model by combining with other advanced neural network-based replay attack detectors. 3) By taking advantage of the fact that replay attack detection does not necessarily need to use the entire sequence, a few strategies such as using part of the input, non-overlap framing, and convolution without padding are used to speed up computation.

\begin{table}[t]
\scriptsize
\centering
\caption{Microphone array settings}
\label{tab:mic_array}
\begin{tabular}{@{}ccccc@{}}
\toprule
Device  &Model             & Sample Rate & Bit Depth & \#Channels \\ \midrule
\textbf{D1}   &Google AIY         & 44100       & 16       & 2          \\
\textbf{D2}   &Respeaker 4 Linear & 44100       & 16       & 4          \\
\textbf{D3}   &Respeaker V2       & 44100       & 32       & 6          \\
\textbf{D4}   &Amlogic 113X1      & 16000       & 16       & 7          \\
\bottomrule
\end{tabular}
\end{table}

\begin{figure}[t]
  \centering
  \includegraphics[width=6.35cm]{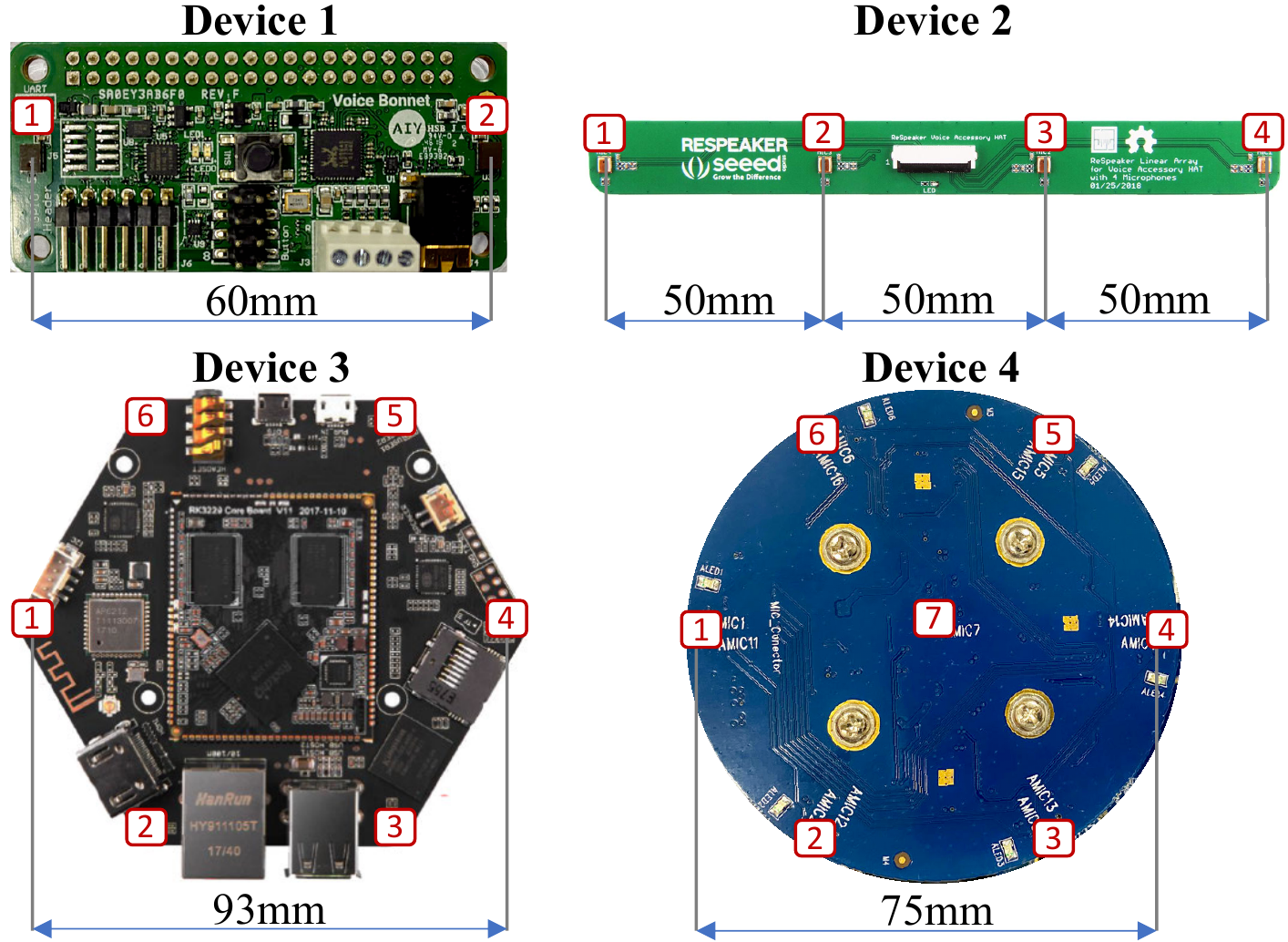}
  \caption{Microphone array geometries used in ReMASC (microphones are shown using the rectangles, where the number indicates the index).}
  \label{fig:mic_array}
\end{figure}

\section{Experimentation}
\subsection{Dataset}

We previously collected the ReMASC (Realistic Replay Attack Microphone Array Speech) corpus~\cite{gong2019remasc} to facilitate experiments on microphone arrays. The ReMASC corpus differs from other publicly available voice anti-spoofing datasets (e.g., the RedDots replayed dataset~\cite{kinnunen2017reddots}) as follows: First, the ReMASC corpus contains recordings by a variety of microphone array systems instead of a single microphone. The microphone array geometry and the corresponding recording settings are shown in Fig.~\ref{fig:mic_array} and Table~\ref{tab:mic_array}, respectively. Second, instead of using audio simulation tools~\cite{sainath2017multichannel}, we recorded the ReMASC corpus in a variety of realistic usage scenarios and settings. Therefore, the ReMASC dataset is particularly well-suited for multi-channel voice anti-spoofing research. More details about the ReMASC corpus can be found in~\cite{gong2019remasc}.

\subsection{Data Splitting and Training Scheme}

The ReMASC corpus is split into the core set (6331 genuine and 17175 replayed samples) and the evaluation set (2118 genuine and 14162 replayed samples). Both sets are speaker-independent and therefore strictly non-overlapping. In this study, we use the core set for training and development, and the evaluation set for testing. Specifically, we use 90\% of the core set to train the model and use the development set to choose the batch size and initial learning rate as well as to implement the early stopping strategy (i.e., stop training when the evaluation metric on the development set stops improving). We use a learning rate decay with warm-up period strategy, i.e., the learning rate starts at the initial learning rate and linearly increases to 10 times larger in the first 20 epochs (the warm-up period), and then drops by half every 20 epochs until the equal error rate (EER) stops improving on the development set or the max epoch of 100 is reached. We select the batch size of 64 and the initial learning rate of 1e-5 through a grid search. We use ADAM optimizer with a l2-norm regularizer~\cite{kingma2014adam,loshchilov2017decoupled} with the weight decay coefficient of 1e-3. All experiments are repeated three times with different random seeds and the mean value is reported. We use EER as our evaluation metric. Since the four microphone arrays were mounted on a stand and recorded simultaneously during the data collection, the data volume recorded by each of the 4 microphone arrays is roughly 1/4 of the total data volume (D1 has fewer data due to hardware crashes in the data collection). Since the microphone arrays use different hardware and geometry, we train and evaluate the machine learning model separately for each microphone array by only using data collected by it.

\subsection{Models}
\label{sec:model}

We compare the following models in our experiments:

\subsubsection{NN-Single} 
This model is exactly the same as the proposed model described in Section~\ref{sec:method} except that only the first channel is fed to the neural network.

\subsubsection{NN-Dummy Multichannel} 
This model is exactly the same as the proposed model in Section~\ref{sec:method}, but here we feed multi-channel input into the neural network. The difference is that we feed replications of the first channel data as ``dummy'' multi-channel input to the neural network. The reason why we include this model in our experiment is that only comparing \emph{NN-Single} with the proposed multi-channel model is not absolutely fair: when the number of input channels increases, the neural network architecture also changes. Specifically, the number of the total front-end filters (i.e., $h$) linearly increases with the input channel numbers. Therefore, it is possible that the performance difference between \emph{NN-Single} and the proposed multi-channel model is actually due to the increasing number of front-end filters instead of the spatial information in the multi-channel input. In contrast, this \emph{NN-Dummy Multichannel} model has exactly the same neural network architecture and the number of parameters as the proposed multi-channel model, and therefore the performance difference can only be attributed to input difference.

\subsubsection{NN-Multichannel} 
This is the proposed model described in Section~\ref{sec:method}. For each microphone array, all channels are used.

\subsection{Results}

\begin{table}[h]
\scriptsize
\centering
\caption{Comparison of EER (\%) of the tested models and the relative improvement of NN-Multichannel over NN-single}
\label{tab:exp1}
\begin{tabular}{@{}lllll@{}}
\toprule
Model                 & D1            & D2            & D3            & D4            \\ \midrule
NN-Single             & 16.6          & 23.7          & 23.7          & 27.5          \\
NN-Dummy Multichannel & 16.0          & 23.2          & 24.5          & 25.2          \\
NN-Multichannel       & \textbf{14.9} & \textbf{15.4} & \textbf{16.5} & \textbf{19.8} \\ \midrule
Multichannel Improvement     & -10.0\% & -34.9\% & -30.3\% & -27.9\% \\ 
\bottomrule
\end{tabular}
\end{table}

\subsubsection{Model Comparison}
\label{sec:exp1}
In this section and in Section \ref{sec:exp2}, we use a fixed $P$ of 64 and the first 1s of each audio sample as the input. As shown in Table~\ref{tab:exp1}, the \emph{NN-Multichannel} model clearly outperforms the \emph{NN-Single} model, with an average EER improvement of 25.8\%. In contrast, the \emph{NN-Dummy Multichannel} performs similarly as \emph{NN-Single}. Since \emph{NN-Multichannel} and \emph{NN-Dummy Multichannel} have exactly the same neural network architecture and the only difference is the input, this demonstrates that the performance improvement of \emph{NN-Multichannel} is not due to its larger number of front-end filters, but due to effectively leveraging the information in the multichannel input. Further, we test the \emph{NN-Single} model using channels other than the first channel. We observe that the performance of the model changes with the used channel (which could be due to some microphones being closer to the speaker for more samples). But the performance variance of the \emph{NN-Single} models using different channels is small compared to the performance difference between \emph{NN-Multichannel} and \emph{NN-Single}. Specifically, the standard deviation of the \emph{NN-Single} performance using different channels is 1.32, 2.01, 1.49, 0.96 (\% EER) for D1, D2, D3, and D4, respectively. Also, we find that the performance of \emph{NN-Multichannel} always outperforms \emph{NN-Single} using any channel by a clear margin, e.g., for D1, \emph{NN-Single} using channel \#1 and channel \#2 achieves EER of 16.6\% and 18.4\%, respectively, but the EER of \emph{NN-Multichannel} is only 14.9\%. Therefore, we believe that the main contributor of the performance improvement of \emph{NN-Multichannel} is that the spatial information is effectively leveraged rather than that there exists a most effective single channel in the multichannel input.

\begin{table}[h]
\scriptsize
\centering
\caption{Impact of the number of input channels}
\label{tab:exp2}
\begin{tabular}{@{}lccccccc@{}}
\toprule
\multicolumn{1}{c}{\multirow{2}{*}{Device}} & \multicolumn{7}{c}{\# Used Channels}                                               \\ \cmidrule(l){2-8} 
\multicolumn{1}{c}{}                        & 1    & 2             & 3    & 4             & 5             & 6             & 7    \\ \midrule
D1                                          & 16.6 & \textbf{14.9} & -    & -             & -             & -             & -    \\
D2 (1-4-2-3)                                & 23.7 & 22.9          & 18.0 & \textbf{15.4} & -             & -             & -    \\
D2 (1-2-3-4)                                & 23.7 & 19.5          & 16.7 & \textbf{15.4} & -             & -             & -    \\
D3                                          & 23.7 & 19.1          & 17.6 & 17.0          & 17.1 & \textbf{16.5}          & -    \\
D4                                          & 27.5 & 21.5          & 20.6 & 21.3          & 20.7          & 19.9 & \textbf{19.8} \\ \bottomrule
\end{tabular}
\end{table}

\subsubsection{Impact of the Number of Input Channels}
\label{sec:exp2}
In this section, we further investigate the impact of the number of input channels. We gradually add the input channels from one channel to the total available channels of each microphone array and measure the EER. The order with which we add the microphones is (microphone indexed as shown in Fig.~\ref{fig:mic_array}): D1: 1-2; D2: 1-4-2-3; D3: 1-4-2-5-3-6; D4: 1-4-2-5-3-6-7. The rule is that we add the microphone furthest from the previously added microphone. For device 2, we further test a different order of 1-2-3-4. As shown in Table~\ref{tab:exp2}, we have the following findings. First, the EER generally drops with an increasing number of used microphones and the best performance of all four devices is achieved when all microphones are used, indicating that more microphones help improve the defense performance. Second, we observe that the performance improves most significantly when the second microphone (the microphone that has the longest distance from the first microphone) is added for D1, D3, and D4, and that the performance gradually saturates with more microphones. Nevertheless, D2 gets the most significant performance improvement when the third microphone is added, and if we switch the order to 1-2-3-4, we find that the most significant performance improvement is achieved when the second microphone is added. This indicates that it is not always optimal to use a microphone array with a larger dimension. We intend to explore the impact of array geometry in our future research.

\begin{table}[htb]
\scriptsize
\centering
\caption{Impact of the number of front-end filters per channel}
\label{tab:exp3}
\begin{tabular}{@{}ccccccc@{}}
\toprule
\multirow{2}{*}{} & \multicolumn{6}{c}{\# Front-end Filters Per Channel} \\ \cmidrule(l){2-7} 
                  & 4        & 8        & 16      & 32      & 64      & 128     \\ \midrule
Mean EER (\%)     & 22.6     & 19.7     & 18.1    & 17.8    & 16.7    & 17.3    \\ \bottomrule
\end{tabular}
\end{table}

\subsubsection{Impact of Other Factors}
\label{sec:exp3}
First, we explore the impact of the number of front-end filters, as shown in Table~\ref{tab:exp3}; the model performance improves with the increasing number of front-end filters until $P=64$. The result is as expected since more filters can help extract spatial and spectral features, but too many parameters make training less effective. Second, we explore the impact of the input audio length. We find that using the first 0.5s, 1s, and 1.5s of the audio leads to a mean EER of 26.9\%, 22.9\%, and 23.4\% for \emph{NN-Single} and 19.6\%, 16.7\%, and 18.9\% for \emph{NN-Multichannel}, respectively. In addition, we find that using a segment from the beginning of the audio slightly outperforms segments from the middle for both \emph{NN-Single} (22.9\% vs 23.8\%) and \emph{NN-Multichannel} (16.7\% vs 17.4\%) when an input audio length of 1s is used. This verifies our discussion from Section II and indicates that using the first 1s of each sample is the most appropriate setting in terms of both performance (for both \emph{NN-Single} and \emph{NN-Multichannel}) and computational overhead.

\section{Conclusions}
In this paper, we introduce a novel neural network-based replay attack detection model that leverages both the spectral and spatial information in the multi-channel audio and is able to significantly improve the replay attack detection performance. Compared to previous efforts, the proposed model supports arbitrary number of input channels and is completely data-driven in a neural network framework, which will make it easy to combine the proposed method with other neural-based anti-spoofing countermeasures. 
%We hope this paper can contribute to future research on multi-channel based voice-antispoofing research.

\bibliographystyle{IEEEtran}
\bibliography{ref}

\end{document}